\documentclass[aps, prd, twocolumn, amsmath, floats,floatfix, superscriptaddress, nofootinbib, showkeys,
showpacs]{revtex4}
\usepackage{amssymb}
\usepackage{amsmath}
\usepackage{verbatim}
\usepackage{mathrsfs}
\usepackage{amsfonts}
\usepackage{latexsym}
\usepackage{epsfig}
\usepackage{color}
\usepackage[usenames,dvipsnames]{xcolor}
\usepackage[colorlinks, linkcolor={Red},citecolor={Blue}]{hyperref}
\usepackage{graphicx,subfigure}
\usepackage{units}
\usepackage{envmath}
\usepackage{natbib}
\usepackage{ctable}
\usepackage{soul}
\begin{document}
\definecolor{orange}{rgb}{0.9,0.45,0}
\def\CovDev{D}
\def\Res{{\mathcal R}}
\def\Gammaflat{\hat \Gamma}
\def\metricflat{\hat \gamma}
\def\Dflat{\hat {\mathcal D}}
\def\part_n{\partial_\perp}
%
\def\Lie{\mathcal{L}}
\def\A{\mathcal{X}}
\def\Aphi{\A_{\phi}}
\def\hAphi{\hat{\A}_{\phi}}
\def\E{\mathcal{E}}
\def\Ham{\mathcal{H}}
\def\M{\mathcal{M}}
\def\R{\mathcal{R}}
\def\p{\partial}
\def\hg{\hat{\gamma}}
\def\hA{\hat{A}}
\def\hD{\hat{D}}
\def\hE{\hat{E}}
\def\hR{\hat{R}}
\def\hcA{\hat{\mathcal{A}}}
\def\hDelt{\hat{\triangle}}
\def\na{\nabla}
\def\dif{{\rm{d}}}
\def\non{\nonumber}
\newcommand{\erf}{\textrm{erf}}
\newcommand{\saeed}[1]{\textcolor{blue}{SF: #1}} 
%
\renewcommand{\t}{\times}
\long\def\symbolfootnote[#1]#2{\begingroup%
\def\thefootnote{\fnsymbol{footnote}}\footnote[#1]{#2}\endgroup}
\title{On the Merger Rate of Primordial Black Holes in Cosmic Voids}

\author{Saeed Fakhry} 
\email{s\_fakhry@sbu.ac.ir}
\affiliation{Department of Physics, Shahid Beheshti University, Evin, Tehran 19839, Iran}
\affiliation{PDAT Laboratory, Department of Physics, K.N. Toosi University of Technology, P.O. Box 15875-4416, Tehran, Iran}

\author{Seyed Sajad Tabasi} 
\email{sstabasi98@gmail.com}
\affiliation{Department of Physics, Sharif University of Technology, P.O. Box 11155-9161, Tehran, Iran}
\affiliation{PDAT Laboratory, Department of Physics, K.N. Toosi University of Technology, P.O. Box 15875-4416, Tehran, Iran}

\author{Javad T. Firouzjaee} 
\email{firouzjaee@kntu.ac.ir}
\affiliation{Department of Physics, K.N. Toosi University of Technology, P.O. Box 15875-4416, Tehran, Iran}
\affiliation{School of Physics, Institute for Research in Fundamental Sciences (IPM), P.O. Box 19395-5531, Tehran, Iran}
\affiliation{PDAT Laboratory, Department of Physics, K.N. Toosi University of Technology, P.O. Box 15875-4416, Tehran, Iran}

\date{\today}
\begin{abstract}
\noindent
Cosmic voids are known as underdense substructures of the cosmic web that cover a large volume of the Universe. It is known that cosmic voids contain a small number of dark matter halos, so the existence of primordial black holes (PBHs) in these secluded regions of the Universe is not unlikely. In this work, we calculate the merger rate of PBHs in dark matter halos structured in cosmic voids and determine their contribution to gravitational wave events resulting from black hole mergers recorded by the Advanced Laser Interferometer Gravitational-Wave Observatory (aLIGO)-Advanced Virgo (aVirgo) detectors. Relying on the PBH scenario, the results of our analysis indicate that about $2\sim 3$ annual events of binary black hole mergers out of all those recorded by the aLIGO-aVirgo detectors should belong to cosmic voids. We also calculate the redshift evolution of the merger rate of PBHs in cosmic voids. The results show that the evolution of the merger rate of PBHs has minimum sensitivity to the redshift changes, which seems reasonable while considering the evolution of cosmic voids. Finally, we specify the behavior of the merger rate of PBHs as a function of their mass and fraction in cosmic voids and we estimate $\mathcal{R}(M_{\rm PBH}, f_{\rm PBH})$ relation, which is well compatible with our findings.

\end{abstract}

\maketitle
\vspace{8cm}

\section{Introduction} 

Primordial black holes (PBHs) are a special type of black holes that form under unique conditions in the earliest evolutionary stages of the Universe \cite{zeld, hawk1, carr+hawk}. When primordial quantum fluctuations exceed a threshold value, they become qualified to directly collapse and lead to the formation of PBHs. The formation of a black hole is itself a fully general relativistic physical process and the evolution of non-linear density perturbations on superhorizon scales and the threshold amplitude of curvature perturbations on superhorizon scales for forming black holes were investigated in several studies \cite{Carr:1975qj, Niemeyer:1999ak, Musco:2004ak, Young:2014ana, musco, Young:2, allah}. In addition, since PBHs form in the early Universe, i.e., before the formation of stars and galaxies, and behave like cosmic fluids at large scales, they can be considered as potential candidates for dark matter depending on their masses \cite{Lacki:2010zf, Belotsky:2014kca, Clesse:2016vqa, Espinosa:2017sgp, Carr:2020xqk, Ashoorioon:2019xqc, Villanueva-Domingo:2021spv}. Moreover, they can provide seeds for the early formation of supermassive black holes (see, e.g., \cite{supermasive-Polnarev:1985btg,supermasive-carr,supermasive-Bean:2002kx}), and are instrumental in the processes occurring in the early Universe, see e.g.,\cite{Green:2014faa,Ashoorioon:2020hln,Carr:2020gox} for a review and references therein.

Theoretically, during the evolution of the Universe, one can expect that PBHs could have been clustered in dark matter halos. Also, due to their random spatial distribution, it makes sense that PBHs in the dark matter halos are capable of encountering each other and/or other compact objects, forming binaries, emitting gravitational waves, and finally merging. After several detections of gravitational waves associated with binary black hole mergers by the Advanced Laser Interferometer Gravitational-Wave Observatory (aLIGO)-Advanced Virgo (aVirgo) detectors \cite{LIGOScientific:2016aoc,LIGOScientific:2021djp,LIGOScientific2021gwtc2}, the interest in PBHs has increased recently.
Most mergers are related to binary black holes with masses of $(10\sim 30)\, M_{\odot}$. In light of this black hole mass range, which often exceeds the astrophysics black hole mass spectrum, it has been suggested that these binary black holes may have primordial origins, see, e.g., \cite{bird, Mandic:2016lcn, Sasaki:2016jop,Kovetz:2017rvv, Sasaki:2018dmp}.

To study the merger rate of PBHs in dark matter halos, two main aspects should be considered. First, the formation mechanism of PBHs, which is related to their contribution to cold dark matter. Second, the physics of dark matter halos within which the PBHs merge. Hence, it can be expected that the concentration parameter of dark matter halos can affect the merger rate of PBHs by modifying their velocity and density distributions. Also, having a statistical view on large scales, it can be found that the mass distribution of dark matter halos can also have a special significance in this case. In this regard, it can be expected that various dark matter halo models provide different predictions of the merger rate of PBHs. Having different types of halo collapse models, the analytically simple one is based on a spherical collapse, which seems unable to provide accurate predictions of local and statistical properties of dark matter halos in some mass limits. To study the role of collapse conditions, in our previous research, we have shown that spherical-collapse dark matter halo models cannot adjust the recorded black hole mergers during the third observing run (O3) in the framework of the PBH scenario, while more realistic halo models (e.g., those with ellipsoidal-collapse) can generate consistent PBH mergers with gravitational wave observations, see our previous studies in Refs.~\cite{Fakhry:2020plg, Fakhry:2021tzk, Fakhry:2022uun, Fakhry:2022hzh} for more details.

On the other side, it is believed that the large-scale visible Universe appears to have a web-like structure called the cosmic web. In some parts of the cosmic web, clusters are separated by large, almost empty regions called cosmic voids where the density of such regions is lower than the average density of the Universe. Being a major part of the cosmic web, cosmic voids are exceptionally underdense regions containing matter, which evacuates them toward other regions. Researches in recent years show that there are galaxies and dark matter halos in cosmic voids \cite{Hahn:2007ui, Sutter:2013mia, Bruton:2019zdo, Tavasoli:2021reo}. According to some theoretical models, dark matter particles are expected to emit detectable gamma-ray signals as a result of their decay and annihilation \cite{Bertone:2007aw, Palomares-Ruiz:2010fpg, Bringmann:2012ez, Blanco:2018esa, Hutten:2022hud}. Additionally, a diffuse background of gamma rays can be detected across the sky by the present gamma-ray observatories \cite{Fermi-LAT:2014ryh}. This background consists of unknown signals that remain after subtracting the contribution from all possible astrophysical sources, such as supermassive black holes and pulsars. Besides, such signals have a non-uniform distribution at different spatial angles, which is adjustable to what is expected from dark matter emission \cite{Karwin:2016tsw}. Accordingly, the way of emitting signals related to the dark matter from overdense and underdense structures of the Universe has been simulated \cite{Arcari:2022zul}. The results indicate that although overall dark matter signals from cosmic voids are weaker, they are less contaminated by astrophysical sources, making them easier to detect. In light of everything discussed so far, the existence of PBHs in cosmic voids is not far from expected. Therefore, it seems interesting to calculate their merger rate in underdense environments with minimal contamination by astrophysical sources.

In this work, we propose to calculate the merger rate of PBHs in the medium of cosmic voids. In this respect, the outline of the work is as follows. In Sec.~\ref{sec:ii}, we briefly discuss cosmic voids and the need to include them in cosmological studies. Then, in Sec.~\ref{sec:iii}, we present a convenient dark matter halo model in cosmic voids and describe some related quantities like halo density profile, halo concentration parameter, and halo mass function. Also, in Sec.~\ref{sec:iv}, we calculate the merger rate of PBHs in cosmic voids, as well as those in other structures. Finally, we discuss the results and summarize the findings in Sec.~\ref{sec:v}.
\section{Cosmic voids} \label{sec:ii} 

The visible Universe on large scales seems to have a web-like structure called the cosmic web. Such a structure is the result of the time evolution of primordial density fluctuations. There are smaller structures inside the cosmic web, including knots, filaments, sheets, and voids, within which matter is distributed differently. Since the initial density field is a Gaussian random field defined by the power spectrum of density fluctuations, and since density fluctuations evolve due to gravity, the gravitational field increases the density contrast in the Universe. As a result, parts of the Universe with stronger gravitational fields become denser over time, while parts with less density become even more empty \cite{Peacock:1993xg}.

In the light these arguments, most of the matter can be distributed in knots, filaments, and sheets. However, clusters are separated by large, almost empty regions called cosmic voids. The density of such regions is lower than the average density of the Univers. Actually, cosmic voids are underdense regions with a low density of matter. According to recent cosmological studies, cosmic voids can provide clues about cosmic mass distribution and serve as a convenient medium for constraining cosmological parameters \cite{Hossen:2021etb, Contarini:2019qwf, Davies:2019yif}. In addition, by taking advantage of the dynamics governing cosmic voids, many studies have been conducted on baryon acoustic oscillations \cite{Khoraminezhad:2021bdl, Zhao:2018hoo}, dark matter-dark energy interaction \cite{Rezaei:2020yhr}, cosmic microwave background \cite{Raghunathan:2019gyb}, and many other hot topics.

In Ref.~\cite{Patiri:2006gr}, a comparison between the properties of various galaxies within cosmic voids has been performed, in which the number density of galaxies is less than $10\%$ of the mean density of the Universe. Also, to find out the properties of cosmic voids, several studies have been carried out via $N$-body simulations \cite{Cautun:2014fwa, Curtis:2021nbi, Baushev:2021hbn}. By considering the size, shape, and structure of cosmic voids, it can be found that such regions might play a prominent role in cosmological evolution because they cover a large volume of the Universe.

Based on the cosmological perturbation theory in the formation and evolution of large-scale structures, a vast majority of matter budget is distributed in dark matter halos \cite{Cooray:2002dia}. Numerical simulations and explanatory studies indicate that the gravitational enhancement of density fluctuations leads to the formation of large-scale structures \cite{Bond:1995yt}. Moreover, a few studies have mapped the population of cosmic voids within the local Universe \cite{Tully:2019ngb, Desmond:2021rih, Shao:2019wit}. Large regions of cosmic voids are related to the foremost noticeable viewpoint of the megaparsec-scale Universe. Cosmic voids are attributed to enormous regions with sizes about $(20\sim 50)\,{\rm Mpc\,h^{-1}}$ that have a lower distribution of matter compared to other structures and possess a significant share of the Universe \cite{Beygu:2016ohd}.

Although some studies have been performed on black holes in cosmic voids, e.g., \cite{Habouzit:2019fij, Capozziello:2004sh, Serpico:2005qz}, no attention has been paid to the merger rate PBHs and their evolution in such regions. As mentioned earlier, it has been suggested that there are galaxies and dark matter halos in cosmic voids. Hence, it is likely that PBHs can be clustered in cosmic voids. Furthermore, research continues on developing new methods for finding galaxies, molecular gas, and star formation in cosmic voids \cite{Das:2015vda, Das:2014pla}. Cosmic voids are interesting from a theoretical perspective because of their huge volume and low density of matter \cite{Voivodic:2020fxt}. In this way, the behavior of dark matter and its candidates can be studied more precisely by considering cosmic voids as solitude regions free of possible astrophysical noises.

Moreover, we might face many challenges when investigating cosmic voids from the standpoints of observation, theory, and simulation. For instance, choosing the right algorithm such as VoidFinder algorithm \cite{Pan:2011hx}, ZOnes Bordering On Voidness (ZOBOV) algorithm \cite{Neyrinck:2007gy}, and DynamIcal Void Analysis (DIVA) algorithm \cite{Lavaux:2009wm} to find cosmic voids is very susceptible. On the other hand, the existence of degeneracies between cosmological parameters in cosmic void simulations cannot be ignored in any way. Regarding this, some studies, e.g., \cite{Sahlen:2015wpc}, have tried to resolve this issue. Furthermore, one must develop some theoretical models that account for galaxy bias and its impact on cosmic void characteristics. However, extensive efforts have been made in this regard \cite{Tinker:2006nn, Little:1993fe, Contarini:2020fdu}.

The structure of cosmic voids is a secluded environment that makes them suitable for observational and gravitational searches. Cosmic voids have less noise than other structures. Thus, it provides a convenient opportunity for focusing on gravitational waves emitted from such regions. Note that the sensitivity of gravitational wave detectors is still not high enough to detect the exact location of merger events. However, with the development of instruments, this can be realized in the upcoming future in such a way that one can witness the classification of merger events arising from the substructures of the cosmic web.
\section{Halo models in cosmic voids}\label{sec:iii}
Dark matter halos are nonlinear cosmological structures that spread in the Universe based on the dynamics governing the formation and evolution of hierarchical structures. They usually originate from physical conditions under which the primordial density fluctuations can be qualified to separate from the expansion of the Universe and collapse due to the self-gravitational force \cite{Ishiyama:2014uoa}. In other words, the physical interpretation of the formation of cosmological structures can be deduced from a dimensionless quantity called density contrast, which is defined as $\delta(r) \equiv [\rho(r)-\bar{\rho}]/\bar{\rho}$. In this relation, $\bar{\rho}$ represents the mean density of the background, and $\rho(r)$ is the density of the overdense region at arbitrary point $r$. According to this definition, which is derived from the excursion sets theory, density fluctuations that exceed the threshold value of overdensities, i.e., $\delta_{\rm c}(z)\simeq 1.686(1+z)$, can provide convenient conditions for the formation of dark matter halos.

On the other hand, the size of cosmic voids and their under-density nature turn them into suitable platforms for investigating the primordial density fluctuations and the formation of cosmological structures such as dark matter halos. Therefore, having proper knowledge about the properties of dark matter halos that form in cosmic voids is necessary \cite{hahnetal2007}. In this regard, the cosmological perturbation theory expresses the density distribution of dark matter particles in galactic halos as a radius-dependent function called the halo density profile. To justify the various fits related to the data obtained from the rotation curve of galaxies, many studies have been carried out to provide a suitable density profile \cite{einasto, jaffe, zeeuw, dehnen, nfw}. One of the most popular density profiles was provided by Navarro, Frenk, and White (NFW), which has the following form \cite{nfw}
\begin{equation}\label{nfw}
\rho(r)=\dfrac{\rho_{\rm s}}{r/r_{\rm s}(1+r/r_{\rm s})^{2}},
\end{equation}
where $\rho_{\rm s}=\rho_{\rm crit}\delta_{\rm c}$ is the scaled density of the halo, $\rho_{\rm crit}$ is the critical density of the background Universe, and $r_{\rm s}$ is the scale radius of the halo. In addition to the NFW profile, the Einasto density profile has been introduced as \cite{einasto}
\begin{equation}\label{einasto}
\rho(r)=\rho_{\rm s}\exp\{-\dfrac{2}{\alpha}\left[\left(\dfrac{r}{r_{\rm s}}\right)^{\alpha}-1\right]\},
\end{equation}
where $\alpha$ is the shape parameter. Although the mentioned density profiles have been derived through different methods, both of them are in good agreement with most of the rotation curve data. In addition, another description of the density profile can be deduced from the concentration parameter, which usually determines the central density of dark matter halos and is defined as
\begin{equation}
C\equiv\dfrac{r_{\rm vir}}{r_{\rm s}},
\end{equation}
where $r_{\rm vir}$ is the halo virial radius. In general, the halo virial radius is attributed to a space that contains a volume whose density is $200$ to $500$ times the critical density of the background Universe. Many attempts have been performed to obtain a suitable concentration parameter \cite{prada, Dutton:2014xda, afshordi, ludlow}. Regarding this, an appropriate analysis that we use in this work is provided in Ref. \cite{afshordi} for the ellipsoidal-collapse dark matter halos.

Moreover, the main challenge regarding the statistics governing dark matter halos is their mass distribution in the cosmic web. Fortunately, in providing a proper statistical description of dark matter halos, a function called the halo mass function can be introduced \cite{Lukic:2007fc}. The halo mass function is a powerful probe in cosmology to classify the mass of dark matter halo structures. In other words, the halo mass function determines the number density of dark matter halos depending on their local quantities. In Ref.~\cite{Jenkins:2000bv}, a suitable description of the scaled differential mass function is presented as 
\begin{equation}
\dfrac{dn}{dM}=g(\sigma)\dfrac{\rho_{\rm m}}{M}\dfrac{d\ln (\sigma^{-1})}{dM},
\end{equation}
where $\rho_{\rm m}$ represents the cosmological matter density, $n(M)$ specifies the number density of dark matter halos with mass $M$, $\sigma(M, z)$ is the variance of linear overdensities on mass $M$ and redshift $z$, and $g(\sigma)$ is the fitting function, which is determined according to the geometrical conditions governing the density field at the collapse time. To obtain a suitable form of the fitting function $g(\sigma)$, several methods have been employed \cite{reed3, warren6, reed6, Tinker:2008ff, ps1974, st1999, smt2001}. One of the most successful functions can be defined as \cite{st1999, smt2001}
\begin{equation}
g_{\rm st}(\sigma)=F\sqrt{\dfrac{2a}{\pi}}\dfrac{\delta_{\rm c}}{\sigma}\exp\left(-\dfrac{a\delta_{c}^{2}}{2\sigma^{2}}\right)\left[1+\left(\dfrac{\sigma^{2}}{a\delta_{\rm c}^{2}}\right)^{p}\right],
\end{equation}
which is known as the Sheth-Tormen (ST) fitting function. In this relation, $F=0.322$, $a=0.707$, and $p=0.3$. It is worth noting that for dark matter halos that are supposed to be virialized in cosmic voids, the matter density parameter can be included in the form of $\Omega_{\rm m, void}=0.03\mbox{-}0.05$ and the dark energy density parameter can be considered as $\Omega_{\Lambda}=0.7$ \cite{Gottloeber:2003zb}. In the next section, we will calculate the merger rate of PBHs in cosmic voids.
\section{PBH merger rate in cosmic voids} \label{sec:iv}
In this section, we aim to calculate the merger rate of those PBHs clustered in dark matter halos in cosmic voids. For this purpose, we have discussed the necessary tools for modeling halos in cosmic voids in the previous section. Assume that two PBHs with masses $m_{1}$ and $m_{2}$ and relative velocity at large separation $v_{\rm r}=|v_{1}-v_{2}|$ accidentally encounter each other in a dark matter halo. Under such assumptions, Keplerian mechanics implies that the maximum gravitational radiation is emitted at the periastron \cite{peters}. By considering the situations under which tidal forces of the surrounding black holes make the head-on collision rare, one can expect that the gravitational interaction between two PBHs yields the formation of a binary system with maximum eccentricity. On the other hand, one can roughly imply that under the strong limits of gravitational focusing, the tidal forces of the surrounding black holes cannot disturb the orbital parameters of the formed binaries, see, e.g., \cite{Fakhry:2020plg, Fakhry:2021tzk, Fakhry:2022hzh} for more information.

Hence, the cross-section for the binary formation can be calculated by the following relation \cite{quinlan, Mouri:2002mc}
\begin{equation}\label{cross-sec}
\chi = 2\pi \left(\dfrac{85\pi}{6\sqrt{2}}\right)^{2/7}\dfrac{G^{2}(m_{1}+m_{2})^{10/7}(m_{1}m_{2})^{2/7}}{c^{10/7}v_{\rm rel}^{18/7}},
\end{equation}
where $G$ is the gravitational constant and $c$ is the velocity of light. We focus on BH binary mergers recorded by the aLIGO-aVirgo detectors, which are traditionally considered to be like $(30\,M_{\odot}\mbox{-}30\,M_{\odot})$ events in galactic halos. For this purpose, we consider the mass of PBHs as $m_{1}=m_{2}=M_{\rm PBH}$ and their relative velocity in the form of $v_{\rm rel}=v_{\rm PBH}$. Therefore Eq.\,\eqref{cross-sec} takes the following form
\begin{multline}
 \chi \simeq 4\pi \left(\frac{85\pi}{3}\right)^{2/7}\left(\frac{M_{\rm PBH}^{2}G^{2}}{c^{10/7}v_{\rm{PBH}}^{18/7}}\right) \simeq 1.37 \times 10^{-14}\times \\ 
 \hspace*{0.5cm}\left(\frac{M_{\rm PBH}}{30\,M_{\odot}}\right)^{2}\left(\frac{v_{\rm PBH}}{200\, \rm km/s}\right)^{-18/7} \rm in\hspace{0.3cm}(pc)^{2},
\end{multline}
where in the second equality, we have normalized the mass of PBHs to $30\,M_{\odot}$ and their velocity to the average velocity of dark matter particles in galactic halos, i.e., $200\,{\rm km/s}$. 

Thus, the merger rate of PBH binaries within each dark matter halo is specified by the following formula \cite{bird}
\begin{equation}\label{eq:per}
 \Upsilon = 2\pi\int_{0}^{r_{\rm vir}} r^{2}\left(\frac{f_{\rm PBH}\,\rho_{\rm h}(r)}{M_{\rm PBH}}\right)^{2}\langle \chi\, v_{\rm PBH}\rangle dr,
\end{equation}
where $\rho_{\rm h}(r)$ represents the halo density profile (see Eqs.\,\eqref{nfw} and \eqref{einasto}), and the angle bracket indicates an average over the PBH relative velocity distribution in the galactic halo. Also, $0<f_{\rm PBH}\,(=\Omega_{\rm PBH}/\Omega_{\rm DM})\leq 1$ is the fraction of PBHs, which determines the contribution of PBHs in dark matter. 

In addition, the halo virial mass, which is considered the mass located within the virial radius is calculated by the following relation
\begin{equation}
M_{\rm vir}=\int_{0}^{r_{\rm vir}}4\pi r^{2}\rho_{\rm h}(r)dr,
\end{equation}
which can be calculated while considering NFW and Einasto density profiles, see \cite{Fakhry:2020plg}.

Another crucial factor for calculating the merger rate of PBHs is the velocity dispersion of dark matter particles in galactic halos. In Ref.~\cite{Gottloeber:2003zb}, it has been argued that the size of the voids is not strongly correlated with the circular velocity limit of the halos used to describe the voids. Hence, in good approximation, it is possible to employ the velocity dispersion relation of dark matter particles obtained in Ref.~\cite{prada} for those halos situated in cosmic voids
\begin{equation}
 v_{\rm disp}=\frac{v_{\rm max}}{\sqrt{2}}=\sqrt{\frac{GM(r<r_{\rm max})}{r_{\rm max}}},
\end{equation}
where $v_{\rm max}$ is the maximum velocity in $r_{\rm max}$ radius. Moreover, we demand that the Maxwell-Boltzmann distribution with a cutoff at the virial velocity handles the relative velocity distributions of PBHs in the galactic halo as the following probability distribution function
\begin{equation}\label{3}
 P(v_{\rm PBH}, v_{\rm disp})=A_{0}\left[\exp\left(-\frac{v_{\rm PBH}^{2}}{v_{\rm disp}^{2}}
 \right)-\exp\left(-\frac{v_{\rm vir}^{2}}{v_{\rm disp}^{2}}\right)\right],
\end{equation}
where $A_{0}$ is characterized by the normalization condition. 

Actually, PBH binary formation can be explained by two different mechanisms, which are not incompatible, but operate in two different epochs of the universe \cite{Sasaki:2018dmp}. In this work, we analyze the formation of PBH binaries in dark matter halos in the late-time Universe. Theoretical predictions from this process suggest that all dark matter could potentially be composed of PBHs \cite{bird, Mandic:2016lcn, Sasaki:2018dmp, Fakhry:2020plg, Fakhry:2021tzk, Fakhry:2022uun, Fakhry:2022hzh}. Nevertheless, the initial clustering may cause PBHs to separate from the Hubble flow and form binaries in the early Universe \cite{Ali-Haimoud:2017rtz, Vaskonen:2019jpv}. Under such conditions, the early universe may contain PBH binaries that emit gravitational waves continuously, gradually shrink, and eventually merge with each other. However, the tidal forces from surrounding PBHs could disrupt some of them before the merger is fulfilled \cite{Kavanagh:2018ggo, Raidal:2018bbj}. Specifically, the merger time of the PBH binaries formed in the early universe is determined by their orbital parameters. On the other hand, the random distribution of PBHs in the early Universe allows them to have various orbital parameters at the time of their binary formation. Hence, there have been some binaries that have already merged, others that are supposed to merge in the present-time Universe, and still others that will merge in the future. Accordingly, LIGO-Virgo observations can be explained by the fact that PBHs make up a very small fraction of dark matter \cite{Hall:2020daa, Hutsi:2020sol, Chen:2021nxo, Franciolini:2022tfm}. However, it is important to note that both mentioned mechanisms are still valid for describing black hole binary merger events and their predictions of the abundance of PBHs are still being validated by the LIGO-Virgo detectors. Here, we consider the contribution of PBHs to dark matter as $f_{\rm PBH}=1$ to calculate the merger rate of PBHs. However, we will discuss the generalization of this assumption to other possible cases, i.e., $f_{\rm PBH}<1$, later.

\begin{figure}[t!]
 \begin{minipage}{1\linewidth}
 \includegraphics[width=1\textwidth]{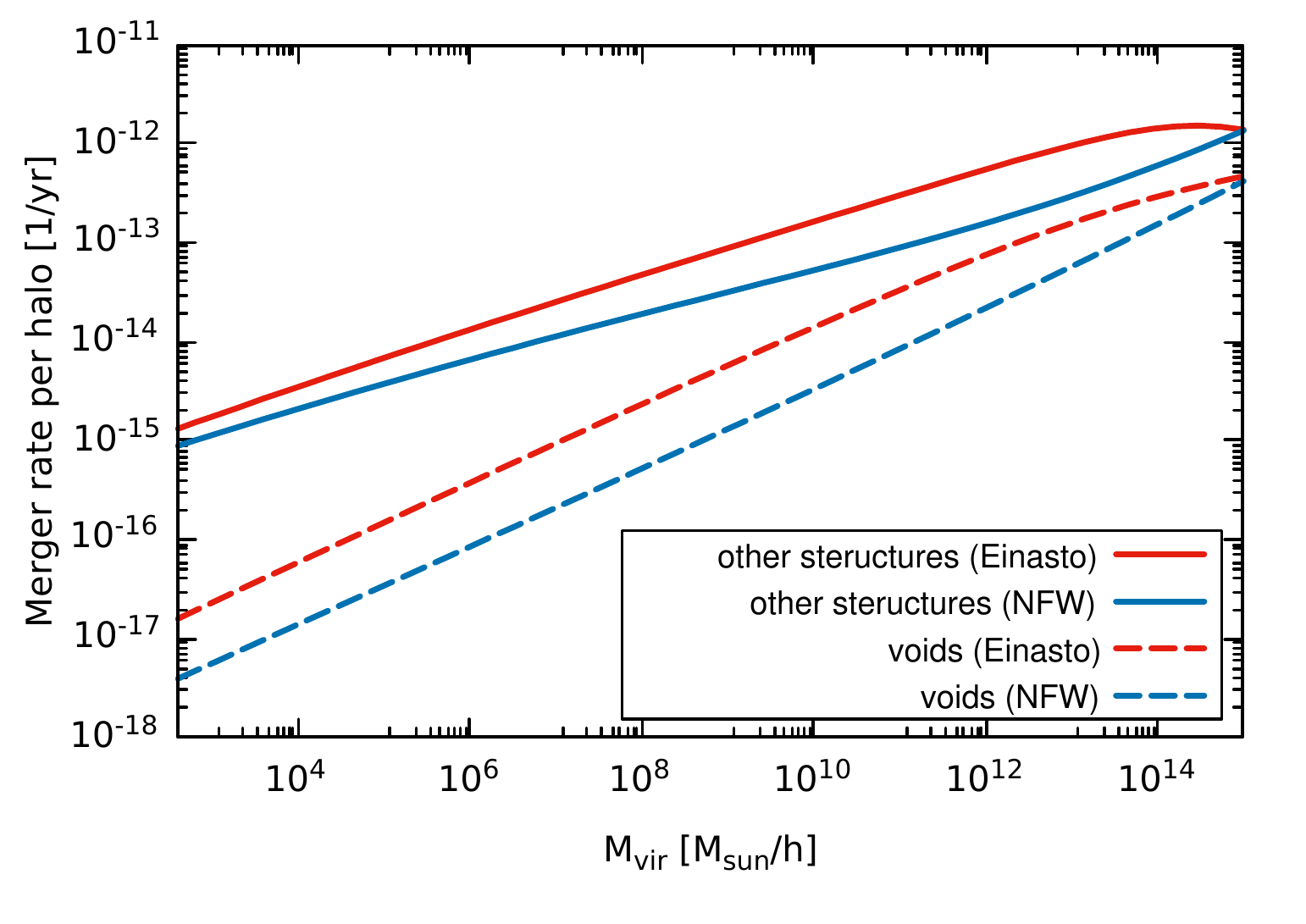}
 \caption{The merger rate of PBHs within each halo as a function of halo mass for NFW and Eniasto density profiles. Solid lines show this relation for dark matter halos in other structures, while dashed lines indicate it for dark matter halos in cosmic voids.}
 \label{fig:per}
 \end{minipage}
\end{figure}

In Fig.~\ref{fig:per}, we have shown the merger rate of PBHs within each halo in cosmic voids and compared it with corresponding results obtained from other structures, while considering NFW and Einasto density profiles. Obviously, the merger rate of PBHs within each halo in cosmic voids is lower than those obtained in halos located in other structures. This result seems reasonable because the number density of halos in cosmic voids is much less than those in other structures. Incidentally, what makes this study interesting comes from the lower number density of the halos, which allows us to take a closer look at the PBH merger rate issue away from all the possible noises in environments with a higher number density of halos. Moreover, our analysis demonstrates that the merger rate of PBHs in halos situated in cosmic voids varies more steeply than those located in other structures. This result confirms the studies conducted on the slope of the primordial power spectrum, which indicates the relative reduction of the number of low-mass halos in voids \cite{WMAP:2003elm}. A thought-provoking point about the formation of halos can be extracted from various numerical analyzes, which show that most of the large mass halos are located in the outer void regions. This could be due to the fact that dark matter halos are not likely to form and cluster in lower-density regions. For this reason, the number density of halos is expected to be much lower in the central void regions than those in outer boundaries.

\begin{figure}[t!]
 \begin{minipage}{1\linewidth}
 \includegraphics[width=1\textwidth]{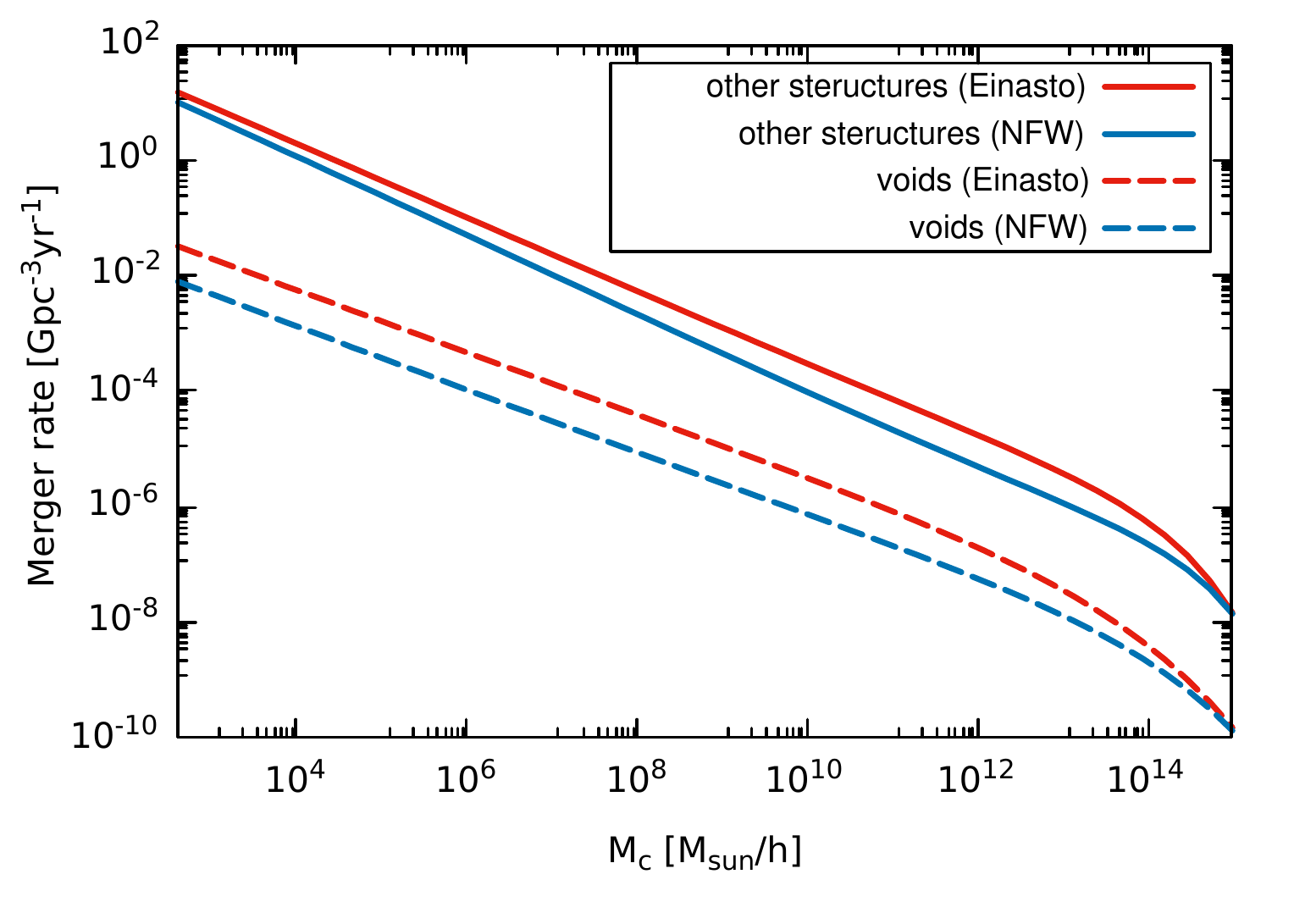}
 \caption{The merger rate of PBHs as a function of halo mass for NFW and Eniasto density profiles. Solid lines demonstrate this relation for dark matter halos in other structures, whereas dashed lines show it for dark matter halos in cosmic voids.}
 \label{fig:tot}
 \end{minipage}
\end{figure}

In addition, what is supposed to be recorded and processed in the aLIGO-aVirgo detectors is the cumulative merger rate of black holes. Hence, the total merger rate of PBHs per unit volume and per unit time must be determined. Regarding this, to calculate the total merger rate of PBHs, one has to convolve the halo mass function, $dn/dM_{\rm vir}$, with the merger rate per halo, $\Upsilon(M_{\rm vir})$:
\begin{equation}\label{tot_mer}
 \mathcal{R}=\int_{M_{\rm c}}\frac{dn}{dM_{\rm vir}} \Upsilon(M_{\rm vir})dM_{\rm vir}.
\end{equation}
It can be easily concluded that the upper limit of integration does not have a significant role in specifying the merger rate of PBHs, because the halo mass function has a decreasing term in such a way that the contribution of the merger rate PBHs decreases exponentially when the mass of halo increases. This is consistent with the hierarchical dynamics of halo formation, as the density of dark matter in low-mass halos is expected to be higher than that in high-mass halos. This is because low-mass halos have been already virialized. As a result, the lower limit of integration plays a vital role in this analysis. According to the argument presented in Refs.~\cite{Fakhry:2020plg, Fakhry:2021tzk}, to calculate the merger rate of PBHs with mass of $30\,M_{\odot}$, one has to consider the lower limit for halo mass to be $M_{\rm c}=400\,M_{\odot}$. This means that signals from dark matter halos with masses less than $400\,M_{\odot}$ are expected to be negligible. Also, in this analysis, we apply the ST halo mass function and the concentration parameter presented in Ref.~\cite{afshordi}.

In Fig.~\ref{fig:tot}, we have depicted the merger rate of PBHs in dark matter halos resided in cosmic voids for NFW and Einasto density profiles and compared it with the results obtained from halos in other structures. Naturally, in this comparison, the contribution of halos in cosmic voids is less than those in other structures. Moreover, as can be seen from the figure, the contribution of small-mass halos to the merger rate of PBHs is still higher than those with larger mass. Because, as mentioned earlier, the density of dark matter particles in subhalos is higher than that in larger halos. As a result, subhalos are expected to be more concentrated than host halos. Note that the total merger rate of PBH binaries can be specified by integrating over the surface below the curves. It can be deduced from this argument that the merger rate of PBHs in dark matter halos located in cosmic voids is of order $(3.4\sim5.9)\times 10^{-2}\,{\rm Gpc^{-3}yr^{-1}}$. At first glance, this value might seem small, but it should be noted that the sensitivity of gravitational wave detectors is such that they are capable to capture merger events up to a comoving volume of $50\,{\rm Gpc^{3}}$ \cite{LIGOScientific2021gwtc2, LIGOScientific:2021djp}. Therefore, based on the PBH scenario, the contribution of dark matter halos in cosmic voids to the merger rate of black holes recorded by gravitational wave detectors is expected to be about $2\sim 3$ events annually.

\begin{figure}[t!]
 \begin{minipage}{1\linewidth}
 \includegraphics[width=1\textwidth]{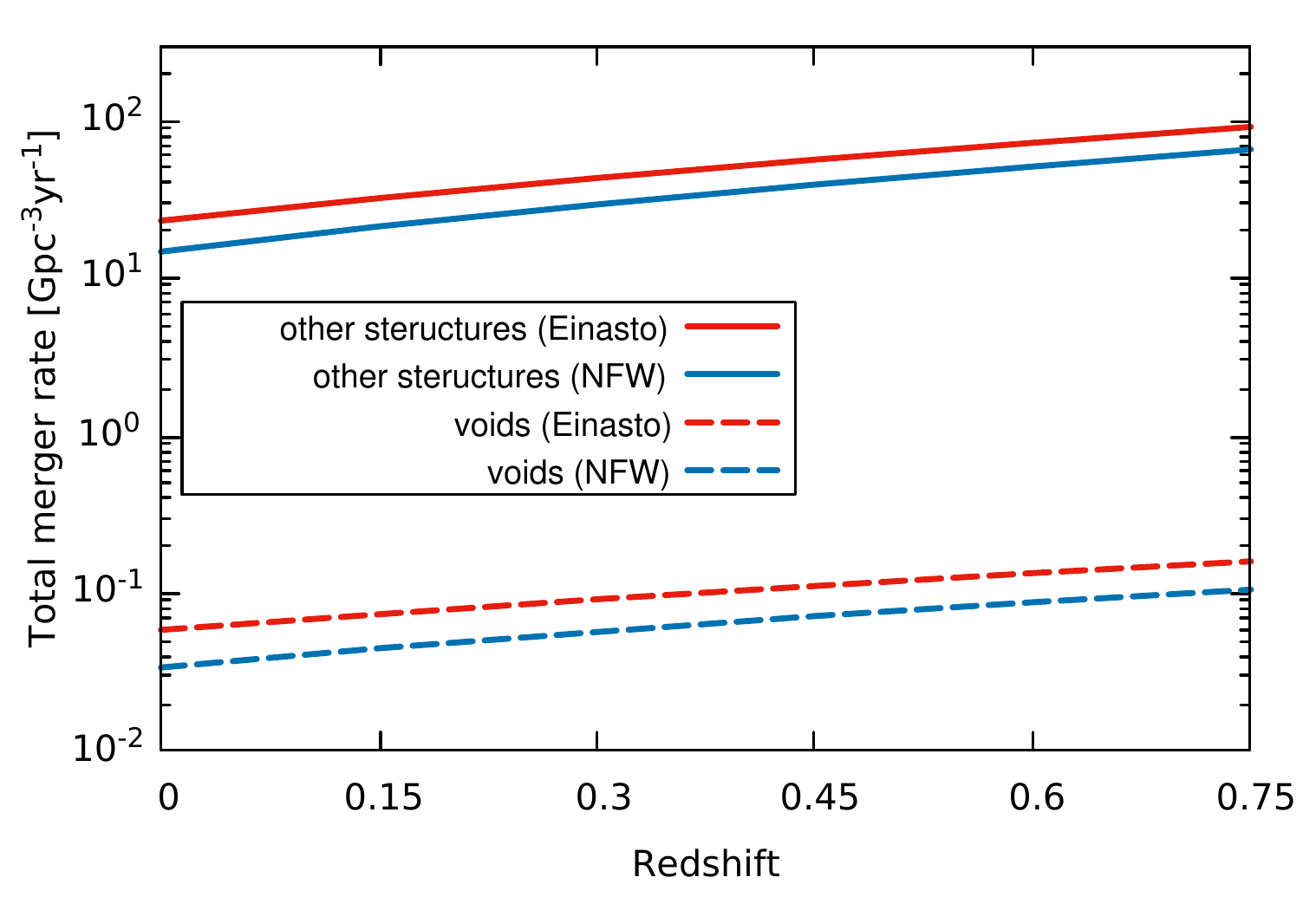}
 \caption{The evolution of the total merger rate of PBHs as a function of redshift for NFW and Eniasto density profiles. Solid lines display this relation for dark matter halos in other structures, whereas dashed lines exhibit it for dark matter halos in cosmic voids.}
 \label{fig:red}
 \end{minipage}
\end{figure}

As mentioned earlier, gravitational wave detectors with their current sensitivity are able to record merger events in the comoving volume $50\,{\rm Gpc^{3}}$, which corresponds to redshift $z\sim 0.75$. Therefore, it is also interesting to calculate the redshift evolution of the merger rate of PBHs. It needs to be reminded that Eq.~\eqref{tot_mer}, through the halo mass function and the concentration parameter, depends on redshift. For this purpose, in Fig.~\ref{fig:red}, we have indicated the redshift evolution of the merger rate of PBHs in halos located in cosmic voids and compared it with the corresponding results for those in other structures. It can be seen that the merger rate of PBHs is directly related to redshift changes. This can be attributed to the fact that relying on the hierarchical dynamics and the halo merger tree, there could have been more subhalos at higher redshifts. Therefore, the merger rate of PBHs in the past was higher than that in the present-time Universe. Also, it can be inferred that such an evolution seems to be common in halos located in cosmic voids and other structures. Another remarkable point can be extracted from the evolution of the merger rate of PBHs in halos located in cosmic voids, which is clearly not very sensitive to redshift changes, while the corresponding results for halos in other structures are strongly sensitive to the redshift changes.

\begin{figure}[t!]
 \begin{minipage}{1\linewidth}
 \includegraphics[width=1\textwidth]{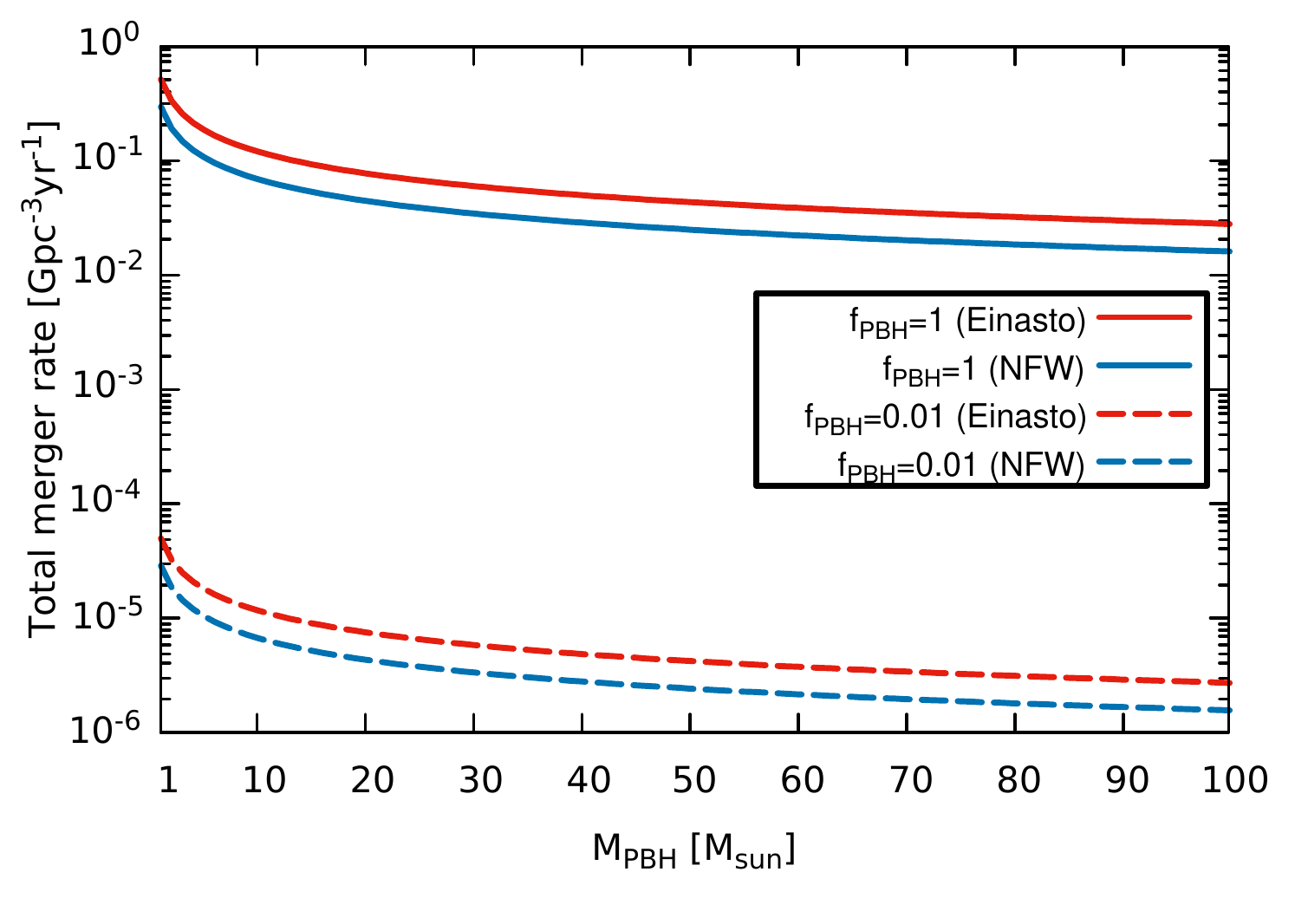}
 \caption{The total merger rate of PBHs as a function of their mass and fraction for NFW and Eniasto density profiles. Solid lines indicate this relation for dark matter halos in other structures, while dashed lines show it for dark matter halos in cosmic voids.}
 \label{fig:frac}
 \end{minipage}
\end{figure}

Up to here, we have carried out the present analysis based on the conditions under which the mass of involving PBHs to be equal to $30\,M_{\odot}$ and their contribution to the dark matter to be in the maximum possible state. Nevertheless, it is possible to calculate the merger rate of PBHs in dark matter halos in the cosmic voids for various masses and fractions of PBHs. For this purpose, we estimate the relationship between the merger rate of PBHs and their masses and fractions as
\begin{equation}\label{tot_mer_es}
\mathcal{R} \simeq \alpha_{\rm void} \, f_{\rm PBH}^{2}\, \left(\dfrac{M_{\rm PBH}}{30\,M_{\odot}}\right)^{-0.63}\hspace*{0.2cm}{\rm in}\hspace*{0.2cm}{\rm (Gpc^{-3}yr^{-1})},
\end{equation}
where $\alpha_{\rm NFW, void}=3.4\times 10^{-2}$ and $\alpha_{\rm Ein, void}=5.9\times 10^{-2}$ are the constants that we have extracted from the relevant calculations for cosmic voids while considering NFW and Einasto profiles. Based on the above relation, in Fig.~\ref{fig:frac}, we have demonstrated the merger rate of PBHs in halos resided in cosmic voids as a function of mass and fraction of PBHs. As can be concluded from the figure and Eq.~\eqref{tot_mer_es}, the merger rate of PBHs changes inversely with their masses. This result seems reasonable since the number density of PBHs is expected to be inversely related to their masses. Therefore, lower-mass PBHs have a higher chance of merging due to their higher number density. In contrast, higher-mass PBHs naturally have a lower merger rate. In addition, we have performed these calculations for two fractions of PBHs, namely $f_{\rm PBH}=0.01$ and $1$. The values corresponding to $0.01<f_{\rm PBH}<1$ fall within this range. Obviously, the direct proportionality of the merger rate of PBHs with their fraction can also be attributed to the number density paradigm. This is because the population of PBHs in a certain volume of dark matter medium corresponds to the description of their number density.

Due to the growing knowledge of gravitational waves, there is a positive outlook on the possibility of recording PBH mergers in cosmic voids by planned gravitational wave detectors. Because one of the main motivations in detecting gravitational waves is to increase the directional precision of networks of gravitational wave detectors. In fact, directional precision measures the average position accuracy of detected signals, and its value directly depends on the size of the detector triangles \cite{Schutz:2011tw}. In other words, it specifies how well the network localizes merger events in the sky. Therefore, extending the baselines in the detection network has a significant effect on increasing the angular accuracy of the transmitted signals. Therefore, by adding ground-based detectors such as Einstein telescope \citep{Maggiore:2019uih} and LIGO-India \cite{Saleem:2021iwi} to the network and launching space-based detectors such as Laser Interferometer Space Antenna (LISA) \cite{Cutler:1997ta} and Deci-hertz Interferometer Gravitational-Wave Observatory (DECIGO) \cite{Kawamura:2020pcg}, it can be possible to localize merger events in the sky. Therefore, the validation of theoretical predictions related to the merger rate of PBHs in cosmic voids using gravitational wave detectors cannot be far from expected.

\section{Conclusions} \label{sec:v}

The cosmic web is one of the most prominent large-scale features of the distribution of dark matter and galaxies in the visible Universe. It includes overdense regions of galaxies, filaments, and sheets, as well as underdense cosmic voids. The main difference between cosmic structures is related to the statistical distribution of matter within them. Meanwhile, cosmic voids comprise a major part of the Universe and are extremely low-dense regions of matter. Although these vast regions of the Universe appear almost empty, cosmic voids have been indicated to contain a distribution of galaxies and dark matter halos. It appears that although dark matter signals from cosmic voids are weak, they are less contaminated by astrophysical sources, making them easier to detect. Under these circumstances, the existence of PBHs, as a potential candidate for dark matter, is not far-fetched in cosmic voids. As a result, calculating the merger rate of PBHs in these regions with the lowest astrophysical noise can lead to an improvement in our knowledge of the formation, evolution, and abundance constraints of PBHs.

In this work, we have calculated the merger rate of PBHs in dark matter halos located in the medium of cosmic voids. For this purpose, we have initially described the theoretical foundations of cosmic voids and their importance in cosmological studies. Containing a large volume of the Universe and accommodating the least amount of matter in themselves, make cosmic voids provide efficient predictions of cosmological consequences. However, the process of these studies might be affected by many observational and theoretical challenges. Furthermore, we have argued that a precise classification of merger events arising from cosmic structures may become possible by increasing the precision of technical instruments. Therefore, the participation of cosmic voids as environments with the lowest amount of astrophysical noises in the PBH merger events can be investigated separately.

In the following, we have structured a theoretical framework for dark matter halo models located in cosmic voids. Dark matter halos are nonlinear hierarchical structures that are formed by initial primordial density fluctuations exceeding a threshold value. Based on this, we have introduced NFW and Einasto density profiles, and the halo concentration parameter, which are expressions of the matter distribution in dark matter halos. We have also explained the halo mass function as a statistical quantity that governs the mass distribution of dark matter halos in cosmic structures.

Having a suitable theoretical model for dark matter halos cosmic voids, we have discussed the encounter conditions of PBHs and their binary formations. Accordingly, we have calculated the merger rate of PBHs within each dark matter halo located in the cosmic voids and presented the corresponding results for those halos in other structures for comparison. The results indicate that the merger rate of PBHs within each halo in cosmic voids is lower than those in other structures, which makes sense due to the incomparable distribution of matter in cosmic voids and other structures. The results also exhibit that the mass variation of the merger rate of PBHs located in cosmic voids is steeper than those located in other structures, which confirms previous studies indicating the reduction of low-mass halos in cosmic voids based on the slope of the primordial power spectrum.

As a consequence, using the main quantities of the presented halo model, we have calculated the cumulative merger rate of PBHs in dark matter halos located in cosmic voids and compared it with the corresponding results obtained from other structures. Our analysis shows that the cumulative merger rate of PBHs present in cosmic voids is also lower than those in other structures. Also, despite the low number density of subhalos in cosmic voids, the results confirm their outstanding contribution to the merger rate of PBHs. This is not surprising because the model of hierarchical structures implies that smaller-mass halos have higher dark matter densities. Our findings suggest that the cumulative merger rate of PBHs in dark matter halos located in cosmic voids is of order $(3.4\sim5.9)\times 10^{-2}\,{\rm Gpc^{-3}yr^{-1}}$, which might seem small at first glance. However, it should be mentioned that the sensitivity of the aLIGO-aVirgo detectors is such that they can record merger events up to a comoving volume of about $50\,{\rm Gpc^{3}}$. Hence, the contribution of dark matter halos in cosmic voids to the merger rate of PBHs recorded by aLIGO-aVirgo detectors is expected to be about $2\sim 3$ events annually.

Relying on the fact that the sensitivity of aLIGO-aVirgo detectors is equivalent to redshift $z\sim 0.75$, we have calculated the redshift evolution of the merger rate of PBHs in cosmic voids as well as other structures. It has been shown that the merger rate of PBHs is directly proportional to redshift changes, which may be attributed to the fact that there could have been more subhalos at higher redshifts. Furthermore, the results demonstrate that the evolution of the merger rate of PBHs in halos located in cosmic voids is not very sensitive to redshift changes, while the corresponding one in other structures is completely sensitive to redshift changes. This can be due to the dynamical evolution of matter in various cosmic structures.

Eventually, we have specified the merger rate of PBHs in dark matter halos located in cosmic voids as a function of mass and fraction of PBHs. The results indicate that the merger rate of PBHs is inversely proportional to their masses.  Also, we have carried out these calculations for two values of the fractions of PBHs as $f_{\rm PBH}=0.01$ and $1$. Particularly, it has been confirmed that the merger rate of PBHs is directly related to their fraction. These results can be due to the fact that the merger rate of PBHs in dark matter halos situated in cosmic voids is strongly related to the number density and mass of PBHs. We have estimated $\mathcal{R}(M_{\rm PBH}, f_{\rm PBH})$ relation, which is well consistent with our findings.


\begin{thebibliography}{99} 

\bibitem{zeld}
Y.B. Zel’dovich and I.D. Novikov,
``The hypothesis of cores retarded during expansion and the hot cosmological model",
{\it Soviet Astron. AJ} (Engl. Transl.) {\bf 10}, 602 (1967).

\bibitem{hawk1}
S. Hawking, 
``Gravitationally collapsed objects of very low mass",
{\it Mon. Not. Roy. Astron. Soc.} {\bf 152}, 75 (1971).

\bibitem{carr+hawk}
B.J. Carr and S. Hawking,
``Black holes in the early universe",
{\it Mon. Not. Roy. Astron. Soc.} {\bf 168}, 399 (1974).

\bibitem{Carr:1975qj}
B.J. Carr,
``The primordial black hole mass spectrum'',
{\it Astrophys. J.}  {\bf 201}, 1 (1975).

\bibitem{Niemeyer:1999ak}
J.C. Niemeyer and K. Jedamzik,
``Dynamics of primordial black hole formation'',
{\it Phys. Rev. D} {\bf 59}, 124013 (1999).

\bibitem{Musco:2004ak}
I. Musco, J.C. Miller and L. Rezzolla,
``Computations of primordial black hole formation'',
{\it Class. Quant. Grav.}  {\bf 22}, 1405 (2005).

\bibitem{Young:2014ana}
A.G. Polnarev and I. Musco,
``Curvature profiles as initial conditions for primordial black hole formation'',
{\it Class. Quant. Grav.}  {\bf 24}, 1405 (2007).

 \bibitem{musco}
I. Musco and J.C. Miller,
``Primordial black hole formation in the early universe: Critical behaviour and self-similarity'',
{\it Class. Quant. Grav.}  {\bf 30}, 145009 (2013).

\bibitem{Young:2}
S. Young, C.T. Byrnes and M. Sasaki,
``Calculating the mass fraction of primordial black holes'',
{\it J. Cosmol. Astropart. Phys.} {\bf 1407}, 045 (2014).

\bibitem{allah}
A. Allahyari, J.T. Firouzjaee and A.A. Abolhasani,
``Primordial black holes in linear and non-linear regimes'',
{\it J. Cosmol. Astropart. Phys.} {\bf 1706}, 041 (2017).

\bibitem{Lacki:2010zf}
B.C. Lacki and J.F. Beacom,
``Primordial black holes as dark matter: Almost all or almost nothing",
{\it Astrophys. J. Lett.} \textbf{720}, L67 (2010).

\bibitem{Belotsky:2014kca}
K.M. Belotsky {\it et al.},
``Signatures of primordial black hole dark matter",
{\it Mod. Phys. Lett. A} \textbf{29}, 1440005 (2014).

\bibitem{Clesse:2016vqa}
S. Clesse and J. Garc\'\i{}a-Bellido,
``The clustering of massive primordial black holes as dark matter: Measuring their mass distribution with advanced LIGO",
{\it Phys. Dark Univ.} \textbf{15}, 142 (2017).

\bibitem{Espinosa:2017sgp}
J.R. Espinosa, D. Racco and A. Riotto,
``Cosmological signature of the standard model Higgs vacuum instability: Primordial black holes as dark matter",
{\it Phys. Rev. Lett.} \textbf{120}, 121301 (2018).

\bibitem{Carr:2020xqk}
B. Carr and F. Kuhnel,
``Primordial black holes as dark matter: Recent developments",
{\it Ann. Rev. Nucl. Part. Sci.} \textbf{70}, 355 (2020).

\bibitem{Ashoorioon:2019xqc}
A. Ashoorioon, A. Rostami and J.T. Firouzjaee,
``EFT compatible PBHs: Effective spawning of the seeds for primordial black holes during inflation",
{\it J. High Energy Phys.} \textbf{07}, 087 (2021).

\bibitem{Villanueva-Domingo:2021spv}
P. Villanueva-Domingo, O. Mena and S. Palomares-Ruiz,
``A brief review on primordial black holes as dark matter",
{\it Front. Astron. Space Sci.} \textbf{8}, 87 (2021).

\bibitem{supermasive-Polnarev:1985btg}
A.G. Polnarev and M.Y. Khlopov,
``Cosmology, primordial black holes, and supermassive particles",
{\it Sov. Phys. Usp.} \textbf{28}, 213 (1985).

\bibitem{supermasive-carr}
B.J. Carr and J.R. Martin,
``Can pregalactic objects generate galaxies?",
{\it Mon. Not. Roy. Astron. Soc.} \textbf{206}, 801 (1984).

\bibitem{supermasive-Bean:2002kx}
R. Bean and J. Magueijo,
``Could supermassive black holes be quintessential primordial black holes?",
{\it Phys. Rev. D} \textbf{66}, 063505 (2002). 

\bibitem{Green:2014faa}
A.M. Green,
``Primordial black holes: Sirens of the early universe",
{\it Fundam. Theor. Phys.} \textbf{178}, 129 (2015).

\bibitem{Ashoorioon:2020hln}
A. Ashoorioon, A. Rostami and J.T. Firouzjaee,
``Examining the end of inflation with primordial black holes mass distribution and gravitational waves",
{\it Phys. Rev. D} \textbf{103}, 123512 (2021).

\bibitem{Carr:2020gox}
B. Carr, K. Kohri, Y. Sendouda and J. Yokoyama,
``Constraints on primordial black holes",
{\it Rept. Prog. Phys.} \textbf{84}, 116902 (2021).

\bibitem{LIGOScientific:2016aoc}
B.P. Abbott \textit{et al.},
``Observation of gravitational waves from a binary black hole merger",
{\it Phys. Rev. Lett.} \textbf{116}, 061102 (2016).

\bibitem{LIGOScientific2021gwtc2}
R. Abbott et al. 
``GWTC-2: Compact binary coalescences observed by LIGO and Virgo during the first half of the third observing run", 
{\it Phys. Rev. X} {\bf 11}, 021053 (2021).

\bibitem{LIGOScientific:2021djp}
R. Abbott \textit{et al.},
``GWTC-3: Compact binary coalescences observed by LIGO and Virgo during the second part of the third observing run",
arXiv:2111.03606.

\bibitem{bird}
S. Bird {\it et al}.
``Did LIGO detect dark matter?'',
{\it Phys. Rev. Lett.} \textbf{116}, 201301 (2016).

\bibitem{Mandic:2016lcn}
V. Mandic, S. Bird and I. Cholis,
``Stochastic gravitational-wave background due to primordial binary black hole mergers",
{\it Phys. Rev. Lett.} \textbf{117}, 201102 (2016).

\bibitem{Sasaki:2016jop}
M. Sasaki, T. Suyama, T. Tanaka and S. Yokoyama,
``Primordial black hole scenario for the gravitational-wave event GW150914'',
{\it Phys. Rev. Lett.} {\bf 117}, 061101 (2016); Erratum:{\it Phys. Rev. Lett.}  {\bf 121}, 059901 (2018).
    
\bibitem{Kovetz:2017rvv}
E.D. Kovetz,
``Probing primordial-black-hole dark matter with gravitational waves",
{\it Phys. Rev. Lett.} \textbf{119}, 131301 (2017).

\bibitem{Sasaki:2018dmp}
M. Sasaki, T. Suyama, T. Tanaka and S. Yokoyama,
``Primordial black holes-perspectives in gravitational wave astronomy",
{\it Class. Quant. Grav.} \textbf{35}, 063001 (2018).

\bibitem{Fakhry:2020plg}
S. Fakhry, J.T. Firouzjaee and M. Farhoudi,
``Primordial black hole merger rate in ellipsoidal-collapse dark matter halo models",
{\it Phys. Rev. D} \textbf{103}, 123014 (2021).

\bibitem{Fakhry:2021tzk}
S. Fakhry, M. Naseri, J.T. Firouzjaee and M. Farhoudi,
``Primordial black hole merger rate in self-interacting dark matter halo models",
{\it Phys. Rev. D} \textbf{105}, 043525 (2022).

\bibitem{Fakhry:2022uun}
S. Fakhry, Z. Salehnia, A. Shirmohammadi and J.T. Firouzjaee,
``The merger rate of primordial black hole-neutron star binaries in ellipsoidal-collapse dark matter halo models",
{\it Astrophys. J.} \textbf{941}, 36 (2022).

\bibitem{Fakhry:2022hzh}
S. Fakhry and A. Del Popolo,
``Effect of a high-precision semianalytical mass function on the merger rate of primordial black holes in dark matter halos",
{\it Phys. Rev. D} \textbf{107}, 063507 (2023).

\bibitem{Hahn:2007ui}
O. Hahn, C.M. Carollo, C. Porciani and A. Dekel,
``The evolution of dark matter halo properties in clusters, filaments, sheets and voids",
{\it Mon. Not. Roy. Astron. Soc.} \textbf{381}, 41 (2007).

\bibitem{Sutter:2013mia}
P.M. Sutter, G. Lavaux, B.D. Wandelt, D.H. Weinberg and M.S. Warren,
``The dark matter of galaxy voids",
{\it Mon. Not. Roy. Astron. Soc.} \textbf{438}, 3177 (2014).

\bibitem{Bruton:2019zdo}
S. Bruton, X. Dai, E. Guerras and F.A. Munshi,
``Deficit of luminous and normal red galaxies in cosmic voids",
{\it Mon. Not. Roy. Astron. Soc.} \textbf{491}, 2496 (2020).

\bibitem{Tavasoli:2021reo}
S. Tavasoli,
``Void galaxy distribution: A challenge for \ensuremath{\Lambda}CDM,''
{\it Astrophys. J. Lett.} \textbf{916}, L24 (2021).

\bibitem{Bertone:2007aw}
G. Bertone, W. Buchmuller, L.Covi and A. Ibarra,
``Gamma-rays from decaying dark matter,",
{\it J. Cosmol. Astropart. Phys} \textbf{11}, 003 (2007).

\bibitem{Palomares-Ruiz:2010fpg}
S. Palomares-Ruiz and J.M. Siegal-Gaskins,
``Annihilation vs. decay: Constraining dark matter properties from a gamma-ray detection",
{\it J. Cosmol. Astropart. Phys} \textbf{07}, 023 (2010).

\bibitem{Bringmann:2012ez}
T. Bringmann and C. Weniger,
``Gamma ray signals from dark matter: Concepts, status and prospects",
{\it Phys. Dark Univ.} \textbf{1}, 194 (2012).

\bibitem{Blanco:2018esa}
C. Blanco and D. Hooper,
``Constraints on decaying dark matter from the isotropic gamma-ray background",
{\it J. Cosmol. Astropart. Phys} \textbf{03}, 019 (2019)

\bibitem{Hutten:2022hud}
M. H\"utten and D. Kerszberg,
``TeV dark matter searches in the extragalactic gamma-ray sky",
{\it Galaxies} \textbf{10}, 92 (2022).

\bibitem{Fermi-LAT:2014ryh}
M. Ackermann \textit{et al.},
``The spectrum of isotropic diffuse gamma-ray emission between 100 MeV and 820 GeV",
{\it Astrophys. J.} \textbf{799}, 86 (2015).

\bibitem{Karwin:2016tsw}
C. Karwin, S. Murgia, T.M.P. Tait, T.A. Porter and P. Tanedo,
``Dark matter interpretation of the Fermi-LAT observation toward the galactic center",
{\it Phys. Rev. D} \textbf{95}, 103005 (2017).

\bibitem{Arcari:2022zul}
S. Arcari, E. Pinetti and N. Fornengo,
``Got plenty of nothing: Cosmic voids as a probe of particle dark matter",
arXiv:2205.03360.

\bibitem{Peacock:1993xg}
J.A. Peacock and S.J. Dodds,
``Reconstructing the linear power spectrum of cosmological mass fluctuations",
{\it Mon. Not. Roy. Astron. Soc.} \textbf{267}, 1020 (1994).

\bibitem{Hossen:2021etb}
M.R. Hossen, S.A. Ema, K. Bolejko and G.F. Lewis,
``Mapping the cosmic mass distribution with stacked weak gravitational lensing and Doppler lensing",
{\it Mon. Not. Roy. Astron. Soc.} \textbf{509}, 5142 (2021).

\bibitem{Contarini:2019qwf}
S. Contarini {\it et al.},
``Cosmological exploitation of the size function of cosmic voids identified in the distribution of biased tracers",
{\it Mon. Not. Roy. Astron. Soc.} \textbf{488}, 3526 (2019).

\bibitem{Davies:2019yif}
C.T. Davies, M. Cautun and B. Li,
``Cosmological test of gravity using weak lensing voids",
{\it Mon. Not. Roy. Astron. Soc.} \textbf{490}, 4907 (2019).

\bibitem{Khoraminezhad:2021bdl}
H. Khoraminezhad, P. Vielzeuf, T. Lazeyras, C. Baccigalupi and M. Viel,
``Cosmic voids and BAO with relative baryon-CDM perturbations",
{\it Mon. Not. Roy. Astron. Soc.} \textbf{511}, 4333 (2022).

\bibitem{Zhao:2018hoo}
C. Zhao {\it et al.},
``Improving baryon acoustic oscillation measurement with the combination of cosmic voids and galaxies",
{\it Mon. Not. Roy. Astron. Soc.} \textbf{491}, 4554 (2020).

\bibitem{Rezaei:2020yhr}
Z. Rezaei,
``Dark matter-dark energy interaction and the shape of cosmic voids",
{\it Astrophys. J.} \textbf{902}, 102 (2020).

\bibitem{Raghunathan:2019gyb}
S. Raghunathan, S. Nadathur, B.D. Sherwin and N. Whitehorn,
``The gravitational lensing signatures of BOSS voids in the cosmic microwave background",
{\it Astrophys. J.} \textbf{890}, 168 (2020).

\bibitem{Patiri:2006gr}
S.G. Patiri, F. Prada, J. Holtzman, A. Klypin and J. Betancort-Rijo,
``The properties of galaxies in voids",
{\it Mon. Not. Roy. Astron. Soc.} \textbf{372}, 1710 (2006).

\bibitem{Cautun:2014fwa}
M. Cautun, R. van de Weygaert, B.J.T. Jones and C.S. Frenk,
``Evolution of the cosmic web",
{\it Mon. Not. Roy. Astron. Soc.} \textbf{441}, 2923 (2014).

\bibitem{Curtis:2021nbi}
O. Curtis, T. Brainerd and A. Hernandez,
``Cosmic voids in GAN-generated maps of large-scale structure",
{\it Astron. Comput.} \textbf{38}, 100525 (2022).

\bibitem{Baushev:2021hbn}
A.N. Baushev,
``The central region of a void: An analytical solution",
{\it Mon. Not. Roy. Astron. Soc.} \textbf{504}, L56 (2021).

\bibitem{Cooray:2002dia}
A. Cooray and R.K. Sheth,
``Halo models of large scale structure",
{\it Phys. Rept.} \textbf{372}, 1 (2002).

\bibitem{Bond:1995yt}
J.R. Bond, L. Kofman and D. Pogosyan,
``How filaments are woven into the cosmic web",
{\it Nature} \textbf{380}, 603 (1996).

\bibitem{Tully:2019ngb}
R.B. Tully {\it et al.},
``Cosmicflows-3: Cosmography of the local void",
{\it Astrophys. J.} \textbf{880}, 24 (2019).

\bibitem{Desmond:2021rih}
H. Desmond, M.L. Hutt, J. Devriendt and A. Slyz,
``Catalogues of voids as antihaloes in the local universe",
{\it Mon. Not. Roy. Astron. Soc.} \textbf{511}, L45 (2022).

\bibitem{Shao:2019wit}
S. Shao, B. Li, M. Cautun, H. Wang and J. Wang,
``Screening maps of the local universe I \textendash{} methodology",
{\it Mon. Not. Roy. Astron. Soc.} \textbf{489}, 4912 (2019).

\bibitem{Beygu:2016ohd}
B. Beygu {\it et al.},
``The void galaxy survey: Star formation properties",
{\it Mon. Not. Roy. Astron. Soc.} \textbf{458}, 394 (2016).

\bibitem{Habouzit:2019fij}
M. Habouzit {\it et al.},
``Properties of simulated galaxies and supermassive black holes in cosmic voids",
{\it Mon. Not. Roy. Astron. Soc.} \textbf{493}, 899 (2020).

\bibitem{Capozziello:2004sh}
S. Capozziello, M. Funaro and C. Stornaiolo,
``Cosmological black holes as seeds of voids in galaxy distribution",
{\it Astron. Astrophys.} \textbf{420}, 847 (2004).

\bibitem{Serpico:2005qz}
M. Serpico, R. D'Abrusco, G. Longo and C. Stornaiolo,
``Cosmological black holes as voids progenitors. 1. Simulations",
{\it Gen. Rel. Grav.} \textbf{39}, 1551 (2007).

\bibitem{Das:2015vda}
M. Das, T. Saito, D. Iono, M. Honey and S. Ramya,
``Detection of molecular gas in void galaxies : Implications for star formation in isolated environments",
{\it Astrophys. J.} \textbf{815}, 40 (2015).

\bibitem{Das:2014pla}
M. Das, T. Saito, D. Iono, M. Honey and S. Ramya,
``Molecular gas and star formation in void galaxies",
{\it IAU Symp.} \textbf{308}, 610 (2014).

\bibitem{Voivodic:2020fxt}
R. Voivodic, H. Rubira and M. Lima,
``The halo void (dust) model of large scale structure",
{\it J. Cosmol. Astropart. Phys} \textbf{10}, 033 (2020).

\bibitem{Pan:2011hx}
D.C. Pan, M.S. Vogeley, F. Hoyle, Y.Y. Choi and C. Park,
``Cosmic voids in sloan digital sky survey data release 7",
{\it Mon. Not. Roy. Astron. Soc.} \textbf{421}, 926 (2012).

\bibitem{Neyrinck:2007gy}
M.C. Neyrinck,
``ZOBOV: A parameter-free void-finding algorithm",
{\it Mon. Not. Roy. Astron. Soc.} \textbf{386}, 2101 (2008).

\bibitem{Lavaux:2009wm}
G. Lavaux and B.D. Wandelt,
``Precision cosmology with voids: Definition, methods, dynamics",
{\it Mon. Not. Roy. Astron. Soc.} \textbf{403}, 1392 (2010).

\bibitem{Sahlen:2015wpc}
M. Sahl\'en, \'I. Zubeld\'\i{}a and J. Silk,
``Cluster-void degeneracy breaking: Dark Energy, Planck, and the largest cluster and void",
{\it Astrophys. J. Lett.} \textbf{820}, L7 (2016).

\bibitem{Tinker:2006nn}
J.L. Tinker, D.H. Weinberg and M.S. Warren,
``Cosmic voids and galaxy bias in the halo occupation framework",
{\it Astrophys. J.} \textbf{647}, 737 (2006).

\bibitem{Little:1993fe}
B. Little and D.H. Weinberg,
``Cosmic voids and biased galaxy formation",
{\it Mon. Not. Roy. Astron. Soc.} \textbf{267}, 605 (1994).

\bibitem{Contarini:2020fdu}
S. Contarini {\it et al.},
``Cosmic voids in modified gravity models with massive neutrinos",
{\it Mon. Not. Roy. Astron. Soc.} \textbf{504}, 5021 (2021).

\bibitem{Ishiyama:2014uoa}
T. Ishiyama,
``Hierarchical formation of dark matter halos and the free streaming scale",
{\it Astrophys. J.} \textbf{788}, 27 (2014).

\bibitem{hahnetal2007}
``Properties of dark matter haloes in clusters, filaments, sheets and voids",
O. Hahn, C. Porciani, C.M. Carollo and A. Dekel,
{\it Mon. Not. Roy. Astron. Soc.} \textbf{375}, 489 (2007).

\bibitem{einasto}
J. Einasto, 
``On the construction of a composite model for the galaxy and on the determination of the system of galactic parameters",
{\it Trudy Astrofizicheskogo Instituta Alma-Ata} \textbf{5}, 87 (1965).

\bibitem{jaffe}
W. Jaffe
``A Simple model for the distribution of light in spherical galaxies",
{\it Mon. Not. Roy. Astron. Soc.} \textbf{202}, 995 (1983).

\bibitem{zeeuw}
T. de Zeeuw,
``Elliptical galaxies with separable potentials",
{\it Mon. Not. Roy. Astron. Soc.} \textbf{216}, 273 (1985).

\bibitem{dehnen}
W. Dehnen,
``A family of potential-density pairs for spherical galaxies and bulges",
{\it Mon. Not. Roy. Astron. Soc.} \textbf{265}, 250 (1993).

\bibitem{nfw}
J.F. Navarro, C.S. Frenk and S.D.M. White,
``The structure of cold dark matter halos",
{\it Astrophys. J.} \textbf{462}, 563 (1996).

\bibitem{prada}
F. Prada \textit{et al}.
``Halo concentrations in the standard $\Lambda$ cold dark matter cosmology",
{\it Mon. Not. Roy. Astron. Soc.} \textbf{423}, 3018 (2012).

\bibitem{Dutton:2014xda}
A.A. Dutton and A.V. Macci\`o,
``Cold dark matter haloes in the Planck era: Evolution of structural parameters for Einasto and NFW profiles'',
{\it Mon. Not. Roy. Astron. Soc.} \textbf{441}, 3359 (2014).

\bibitem{afshordi}
C. Okoli and N. Afshordi,
``Concentration, ellipsoidal collapse, and the densest dark matter haloes'',
{\it Mon. Not. Roy. Astron. Soc.} \textbf{456}, 3068 (2016). 
    
\bibitem{ludlow}
A.D. Ludlow {\it et al}.
``The mass-concentration-redshift relation of cold and warm dark matter haloes'',
{\it Mon. Not. Roy. Astron. Soc.} \textbf{460}, 1214 (2016).

\bibitem{Lukic:2007fc}
Z. Lukic, K. Heitmann, S. Habib, S. Bashinsky and P.M. Ricker,
``The halo mass function: High redshift evolution and universality'',
{\it Astrophys. J.} \textbf{671}, 1160 (2007).

\bibitem{Jenkins:2000bv}
A. Jenkins {\it et al}.
``The mass function of dark matter halos'',
{\it Mon. Not. Roy. Astron. Soc.} \textbf{321}, 372 (2001). 

\bibitem{reed3} 
D. Reed {\it et al.},
``Evolution of the mass function of dark matter haloes",
{\it Mon. Not. Roy. Astron. Soc.} {\bf 346}, 565 (2003).

\bibitem{warren6}
M.S. Warren, K. Abazajian, D.E. Holz and L. Teodoro,
``Precision determination of the mass function of dark matter halos",
{\it Astrophys. J.} {\bf 646}, 881 (2006).

\bibitem{reed6}
D. Reed, R. Bower, C. Frenk, A. Jenkins and T. Theuns,
``The halo mass function from the dark ages through the present day",
{\it Mon. Not. Roy. Astron. Soc.} {\bf 374}, 2 (2007).

\bibitem{Tinker:2008ff}
J.L. Tinker {\it et al.},
``Toward a halo mass function for precision cosmology: The Limits of universality",
{\it Astrophys. J.} \textbf{688}, 709 (2008).

\bibitem{ps1974}
W.H. Press and P. Schechter,
``Formation of galaxies and clusters of galaxies by self-similar gravitational condensation",
{\it Astrophys. J.} {\bf 187}, 425 (1974).

\bibitem{st1999}
R.K. Sheth and G. Tormen,
``Large scale bias and the peak background split",
{\it Mon. Not. Roy. Astron. Soc.} {\bf 308}, 119 (1999). 

\bibitem{smt2001}
R.K. Sheth, H.J. Mo and G. Tormen,
``Ellipsoidal collapse and an improved model for the number and spatial distribution of dark matter haloes", 
{\it Mon. Not. Roy. Astron. Soc.} {\bf 323}, 1 (2001).

\bibitem{Gottloeber:2003zb}
S. Gottloeber, E.L. Lokas, A. Klypin and Y. Hoffman,
``The structure of voids",
{\it Mon. Not. Roy. Astron. Soc.} \textbf{344}, 715 (2003).

\bibitem{peters}
P.C. Peters,
``Gravitational radiation and the motion of two point masses",
{\it Phys. Rev.} {\bf 136}, B1224 (1964).

\bibitem{quinlan}
G.D. Quinlan, S.L. Shapiro,
``Dynamical evolution of dense clusters of compact stars",
{\it Astrophys. J.} \textbf{343}, 725 (1989).

\bibitem{Mouri:2002mc}
H. Mouri and Y. Taniguchi,
``Runaway merging of black holes: Analytical constraint on the timescale'',
{\it Astrophys. J. Lett.} \textbf{566}, L17 (2002).
    
\bibitem{Ali-Haimoud:2017rtz}
Y. Ali-Ha\"\i{}moud, E.D. Kovetz and M. Kamionkowski,
``Merger rate of primordial black-hole binaries",
{\it Phys. Rev. D} \textbf{96}, 123523 (2017).

\bibitem{Vaskonen:2019jpv}
V. Vaskonen and H. Veerm\"ae,
``Lower bound on the primordial black hole merger rate",
{\it Phys. Rev. D} \textbf{101}, 043015 (2020). 
  
\bibitem{Kavanagh:2018ggo}
B.J. Kavanagh, D. Gaggero and G. Bertone,
``Merger rate of a subdominant population of primordial black holes",
{\it Phys. Rev. D} \textbf{98}, 023536 (2018).

\bibitem{Raidal:2018bbj}
M. Raidal, C. Spethmann, V. Vaskonen and H. Veerm\"ae,
``Formation and evolution of primordial black hole binaries in the early universe",
{\it J. Cosmol. Astropart. Phys.} \textbf{02}, 018 (2019).

\bibitem{Hall:2020daa}
A. Hall, A.D. Gow and C.T. Byrnes,
``Bayesian analysis of LIGO-Virgo mergers: Primordial vs. astrophysical black hole populations",
{\it Phys. Rev. D} \textbf{102}, 123524 (2020).

\bibitem{Hutsi:2020sol}
G. H\"utsi, M. Raidal, V. Vaskonen and H. Veerm\"ae,
``Two populations of LIGO-Virgo black holes",
{\it J. Cosmol. Astropart. Phys.} \textbf{03}, 068 (2021).

\bibitem{Chen:2021nxo}
Z.C. Chen, C. Yuan and Q.G. Huang,
``Confronting the primordial black hole scenario with the gravitational-wave events detected by LIGO-Virgo",
{\it Phys. Lett. B} \textbf{829}, 137040 (2022).

\bibitem{Franciolini:2022tfm}
G. Franciolini, I. Musco, P. Pani and A. Urbano,
``From inflation to black hole mergers and back again: Gravitational-wave data-driven constraints on inflationary scenarios with a first-principle model of primordial black holes across the QCD epoch",
{\it Phys. Rev. D} \textbf{106}, 123526 (2022).

\bibitem{WMAP:2003elm}
D.N. Spergel \textit{et al.},
``First year Wilkinson microwave anisotropy probe (WMAP) observations: Determination of cosmological parameters",
{\it Astrophys. J. Suppl.} \textbf{148}, 175 (2003).

\bibitem{Schutz:2011tw}
B.F. Schutz,
``Networks of gravitational wave detectors and three figures of merit",
{\it Class. Quant. Grav.} \textbf{28}, 125023 (2011).

\bibitem{Maggiore:2019uih}
M. Maggiore \textit{et al.},
``Science case for the Einstein telescope",
{\it J. Cosmol. Astropart. Phys.} \textbf{03}, 050 (2020).

\bibitem{Saleem:2021iwi}
M. Saleem \textit{et al.},
``The science case for LIGO-India",
{\it Class. Quant. Grav.} \textbf{39}, 025004 (2022).

\bibitem{Cutler:1997ta}
C. Cutler,
``Angular resolution of the LISA gravitational wave detector",
{\it Phys. Rev. D} \textbf{57}, 7089 (1998).

\bibitem{Kawamura:2020pcg}
S. Kawamura \textit{et al.},
``Current status of space gravitational wave antenna DECIGO and B-DECIGO",
{\it Prog. Theor. Exp. Phys.} \textbf{2021}, 05A105 (2021).

\end{thebibliography}
\end{document}